\documentclass[10pt]{elsart}

\usepackage{epsfig}
\usepackage{subfigure}

\usepackage{amssymb}

\begin{document}

\begin{frontmatter}



\title{The Social Architecture of Capitalism}
\thanks{Thanks to Moshe Machover, Emmanuel Farjoun, Julian Wells, John Funge, Shariffa Karimjee and members of the OPE-L discussion list for comments and criticisms.}
\author{Ian Wright}
\ead{wright@ikuni.com}
\ead[url]{ianusa.home.mindspring.com}
\address{iKuni Inc.,\\ 3400 Hillview Avenue, Building 5, Palo Alto, CA 94304, USA\\Fax: +1 650 320 9827\\Phone: +1 650 739 5355}

\title{}

\begin{center}
{\small
Version 1 December 2003 \\
Version 2 January 2004
}
\end{center}


\author{}

\address{}

\begin{abstract}
A dynamic model of the social relations between workers and
capitalists is introduced. The model is deduced from the assumption
that the law of value is an organising principle of modern economies.
The model self-organises into a dynamic equilibrium with 
statistical properties that are in close qualitative and in 
many cases quantitative agreement with a broad range of known empirical 
distributions of developed capitalism, including the power-law distribution
of firm size, the Laplace distribution of firm and GDP growth, the lognormal 
distribution of firm demises, the exponential distribution of the duration 
of recessions, the lognormal-Pareto distribution of income, 
and the gamma-like distribution of the rate-of-profit of firms. Normally these 
distributions are studied in isolation, but this model unifies and
connects them within a single causal framework. In addition, the model generates
business cycle phenomena, including fluctuating wage and profit shares
in national income about values consistent with empirical studies.
A testable consequence of the model is a conjecture that the
rate-of-profit distribution is consistent with a parameter-mix
of a ratio of normal variates with means and variances that depend on
a firm size parameter that is distributed according to a power-law.
\end{abstract}

\end{frontmatter}


\section{Introduction}

The dominant social relation of production within capitalism is
that between capitalists and workers. A small class of capitalists
employ a large class of workers organized within firms of various
sizes that produce goods and services for sale in the marketplace.
Capitalist owners of firms receive revenue and workers receive a
proportion of the revenue as wages.

Over the last hundred years or more the number and type of material
objects and services processed by capitalist economies have significantly
changed, but the social relations of production have not.
Marx \cite{marx1} proposed the distinction between the forces of production and
the social relations of production to convey this idea.
The existence of a social relationship between a class of capitalists
and a class of workers mediated by wages and profits is an
invariant feature of capitalism, whereas the types of objects and
activities subsumed under this social relationship is not.

The social relations of production constitute an abstract,
but nevertheless real, enduring social architecture that
constrains and enables the space of possible economic interactions. These
social constraints are distinct from any natural or technical
constraints, such as those due to scarcities or current production
techniques. Therefore, unlike many economic models that theorise
relations of utility between economic actors and
scarce commodity types (i.e., actor to object relations studied
under the rubric of neo-classical economics), or theorise relations of
technical dependence between material inputs and outputs (i.e., object to
object relations studied under the rubric of neo-Ricardian economics)
the model developed here entirely abstracts from these relations
and instead theorises relations of social dependence mediated by
value (i.e., actor to actor relations, arguably a defining feature of
Marxist economics). The model ontology is therefore
quite sparse, consisting solely of actors and money. The aim
is to concentrate as far as possible on the economic consequences
of the social relations of production alone, that is on the
enduring social architecture, rather than particular and perhaps
transitory economic mechanisms, such as particular markets, commodity
types and industries. As the worker-capitalist relation is dominant
in developed capitalism the model abstracts from land, rent, states
and banking. Plus, there are important causal relationships between
the forces and relations of production, but in this paper they are
ignored.

In what follows a dynamic, computational model of the social architecture
of capitalism is defined and a preliminary analysis given. It replicates
some important empirical features of modern capitalism. 

\section{A dynamic model of the social relations of production}

The elements of the model are a set of $N$ economic actors
(labelled $1, \ldots, N$). Each actor $i$ at time $t$ holds a non-negative
endowment of paper money, $m_{i}(t) \in \Rset^{+}$, measured in `coins'.
Mechanisms that alter the money supply are ignored and therefore the total
money in the economy is a finite constant $M$, such that
$\sum_{i=1}^{N} m_{i}(t) = M$.
Each actor $i$ has an integer {\em employer index}, $0 \leq e_{i}(t) \leq N$
and $e_{i}(t) \neq i$, which specifies the actor's employer.
If $e_{i}(t) = 0$ the actor is not employed,
otherwise if $e_{i}(t) = j$ then actor $j$ is the employer of actor $i$.
The state of each actor is therefore fully
specified by the pair $(m_{i}, e_{i})$, and the static state of the
whole economy at time $t$ is simply the set of pairs
$S(t) = \{ (m_{i}, e_{i}) : 1 \leq i \leq N \}$.

The actors in the economy are naturally partitioned into three
sets or classes: an employing or capitalist class $C$,
an employee or working class $W$, and an unemployed
class $U$. An actor is a capitalist, $i \in C$, if its {\em employee set} is
non-empty, $W_{i}(t) = \{ x : e_{x} = i, 1 \leq x \leq N \} \neq
\emptyset$. An actor is a worker, $i \in W$, if its employee index is non-zero,
$e_{i} \neq 0$. An actor is unemployed, $i \in U$, if its employee index is
zero, $e_{i} = 0$. In this model an actor cannot belong to
more than one class; therefore the sets of capitalists, workers
and unemployed are disjoint, $C \cap W \cap U = \emptyset$,
and their union forms the set of all actors, $A = C \cup W \cup U$.
More formally, $C = \{x : W_{x} \neq \emptyset \wedge e_{x} = 0\}$,
$W = \{ x : W_{x} = \emptyset \wedge e_{x} \neq 0 \}$ and
$U = \{ x : W_{x} = \emptyset \wedge e_{x} = 0 \}$. The structure
of a firm is simply an employer and set of employees, and firm
ownership is limited to a single capitalist employer: there are
no stocks or joint ownership. Although the total number of actors is fixed this can be
interpreted as a stable workforce in which individuals enter and
exit the workforce at the same rate. An actor, therefore,
represents an abstract role in the economy, rather than a
specific individual. The state evolution of the economy, $S_{t} \rightarrow S_{t+1}$,
is determined by a set of predominately stochastic transition rules,
which are applied at each time step.
Processes that involve subjective indeterminacy (e.g., deciding to
act in a given period) or elements of chance (e.g., finding a buyer
in the marketplace) are modelled by selection from a bounded set
according to a given probability distribution. Often the chosen
distribution is uniform in accordance with Bernoulli's Principle of Insufficient
Reason, which states that in the absence of knowledge to the contrary assume
all outcomes are equally likely. More generally, each employment of a
uniform distribution can be considered as a default functional parameter 
of the model, which may be replaced with a different distribution that
has empirical support.

\subsection{The active actor}

Each actor in the economy performs actions on average at the same rate,
which is modelled by allowing each actor an equal chance to perform its
actions in a given time period. Note however that an actor may act
multiple times in a given period, or not at all. The following rule
selects an {\em active actor} who subsequently has the opportunity to
perform economic actions. The unit of time is interpreted as a
single month of real time, and therefore each actor is active on average
once each month.
\begin{quotation}
\underline{Actor selection rule ${\bf S_{1}}$}: (Stochastic).
\begin{enumerate}
\item Randomly select an actor $a$ from the set $A$ according to a uniform
distribution.
\end{enumerate}
\end{quotation}

\subsection{Employee hiring}

The labour market is modelled in a simple manner.
All unemployed actors seek employment, and all employers hire if they
have sufficient {\it ex ante} funds to pay the average wage.
The wage interval, $\omega = [w_{a},w_{b}]$, is a fixed,
exogenous parameter to the model. Wages are randomly chosen from the
wage interval according to a uniform distribution; hence the average wage is
$\langle w \rangle = (w_{a} + w_{b})/2$. The following hiring rule
is applied if the active actor selected from rule ${\bf S_{1}}$ is unemployed.
\begin{quotation}
\underline{Hiring rule ${\bf H_{1}}(a)$}: (Stochastic).
\begin{enumerate}
\item If actor $a$ is unemployed, $a \in U$, then:
\begin{enumerate}
\item Form the set of potential employers,
$H=\{x: e_{x} = 0, 1 \leq x \leq N \} = C \cup U$.
\item Select an employer, $c \in H$, according to the probability function:
\begin{equation}
P(c)= \frac{m_{c}}{\sum_{x \in H} m_{x}}
\end{equation}
that weighs potential employers by their wealth.
\item If $c$'s money endowment $m_{c}$ exceeds the average wage,
$m_{c} > \langle w \rangle$, then $c$ hires $a$ (set $e_{a} = c$).
\end{enumerate}
\end{enumerate}
\end{quotation}
Hiring rule ${\bf H_{1}}$ allows all non-workers to potentially
hire employees, including hiring by other unemployed individuals
to form new firms, but the chances of hiring favour those employers
with greater wealth, a stochastic bias that represents
the tendency of firm growth to depend on accumulation of capital
out of current profits \cite{kalecki54}. But the stochastic nature
of the rule reflects the innumerable concrete reasons
why particular firms are willing and able to hire more workers
than others.

\subsection{Expenditure on goods and services}

Each actor spends its income on goods and services produced by firms.
But the particular purchases of an individual agent
are not modelled. Instead, they are aggregated into a single
amount that represents the actor's total expenditure for the month.
The total expenditure can represent multiple small purchases,
a single large purchase, or a fraction of a purchase amortized over
several months: the interpretation is deliberately flexible.
Absent a theory of consumption patterns the only relevant
information is that expenditure is bound by the amount of money
an actor has. For simplicity assume that the amount spent
is bound by the actor's coin endowment on a randomly selected day.
A {\em consumer actor} is selected to spend its income but
the spent income is not immediately transferred to firms. Instead,
it is added to a pool of market value that represents the currently
available sum of consumer expenditures, which firms compete for.
\begin{quotation}
\underline{Expenditure rule ${\bf E_{1}}(a)$}: (Stochastic).
\begin{enumerate}
\item Randomly select a consumer actor $b$ from the set $A - \{ a\}$ according to
a uniform distribution.
\item Randomly select an expenditure amount, $m$, from the interval
$[0, m_{b}]$ according to a uniform distribution.
\item Add the $m$ coins to the available market value, $V$. (Hence,
$m_{b}$ is reduced by $m$ and $V$ is increased by $m$.)
\end{enumerate}
\end{quotation}
This rule controls the expenditure of all consumers, whether
workers, capitalists or unemployed. Different classes
spend for different reasons, in particular workers normally
spend their incomes on consumption goods, whereas capitalists
not only consume but invest. The payment of wages is treated
separately, and therefore capitalist expenditure according to
rule ${\bf E_{1}}$ is interpreted as expenditure on
non-wage goods, such as capital goods or personal consumption.
The expenditure rule is also implicitly a saving rule as in
a given period the probability of an actor spending all its
wealth is low.

\subsection{Interaction between firms and the market}

To simplify matters it is assumed that all means of production are
controlled by capitalist owners and therefore individual actors
are unable to produce. Self-employment is ignored in this model:
productive work resulting in saleable goods or services is
performed only by actors within firms.

Each firm produces some
collection of use-values that it attempts to sell in the marketplace.
But individual commodity types and sales are not modelled. Instead,
the total volume of a firm's sales in a given period are disaggregated
into {\em market samples}, which are transfers of money from marketplace
to seller, representing multiple separate transactions, or fractions of
a single large transaction. At this level of abstraction the mapping
from market samples to actual material exchanges is ignored and assumed
to be arbitrary.

Under normal circumstances a firm expects that a worker's labour
adds a value to the product that is bound from below by the wage.
A firm's markup on costs reflects this value expectation, which
may or may not be validated in the market. Obviously, there
are multiple and particular reasons why a worker adds more or less
value to the firm's total product, most of which are difficult to
measure, as partially reflected in the large variety of contested
and negotiable compensation schemes. The relation between concrete
labour and value-added is modelled by assuming that a firm randomly
samples the market once for every employee, to reflect the fact that
each worker potentially adds value, but random to reflect
contingency and subsume the range of possibilities,
from slackers to Stackhonovites, or from replaceable administrators
to irreplaceable film stars. This is a weak formulation of the law
of value \cite{rubin,marx1,wright03}, which implies that, absent
profit-equalizing mechanisms and rents, there is a statistical
tendency for the value of a firm's product to be
linearly related to the amount of social labour-time expended on the
product. In this formulation of the model there is no constant capital, 
and therefore issues arising from the divergence of prices from labour 
values are ignored (see refs. \cite{shaikh84,shaikh98} for a discussion).

Each firm therefore samples the market to gain revenue for every
worker employed. In an idealised  freely competitive economy there is a tendency for
particular production advantages to be regularly adopted by
competing firms, including the removal of scarcities due to employment of
particular kinds of skilled labour. Therefore, the determinants of
the value-added per worker may be considered statistically uniform
across firms.
The statistical variation is interpreted as representing transient differences
in the value creating property of different concrete labours.
The value-added by an active worker to the firm's product
is represented by a transfer of money from the current available
market value $V$. The actual value received in money-form
depends on the prevailing market conditions, and mismatches between value
and exchange-value, or more plainly, costs and revenue, determine
whether firms are rewarded with profits for performing socially-necessary
labour.

The revenue received from the market is the legal property of
the capitalist owner. Capitalist owners therefore accrue revenue via market samples
that represent the social utility of the efforts of their workers.
All these abstractions are expressed in the following market
sample rule:
\begin{quotation}
\underline{Market sample rule ${\bf M_{1}}(a)$}: (Stochastic).
\begin{enumerate}
\item If $a$ is not unemployed, $a \ni U$, then:
\begin{enumerate}
\item Randomly select a revenue amount $m$ from the interval
$[0, V]$ according to a uniform distribution ($V$ is reduced
by $m$.)
\item If actor $a$ is an employee, $a \in W$, then transfer
$m$ coins to the employer, actor $e_{a} \in C$
(hence $m_{e_{a}}$ is increased by $m$.) Alternatively,
if actor $a$ is a capitalist owner, $a \in C$, then transfer
$m$ coins to actor $a$ (hence $m_{a}$ is increased by $m$).
In either case the transferred coins are counted as firm revenue.
\end{enumerate}
\end{enumerate}
\end{quotation}

The money received may represent value embodied in many different
kinds of products and services that are sold in arbitrary amounts to
arbitrary numbers of buyers. The market sample rule abstracts from
the details of individual market transactions and may be interpreted
as modelling the aggregate effect of a dynamic random graph that 
links sellers to buyers in each market period.
The stochastic nature of the rule subsumes innumerable reasons why
particular firms enjoy particular revenues: the only constraints are that
revenue received is determined by the available value in the
marketplace, and that a firm with more employees will on average sample
the market on more occasions than a firm with fewer employees, a bias
justified by the law of value.

\subsection{Employee firing}

If the revenue received by a firm is insufficient to pay the wage bill then
the employer must reduce costs and fire employees. This is captured
by the following firing rule:
\begin{quotation}
\underline{Firing rule ${\bf F_{1}}(a)$}: (Deterministic).
\begin{enumerate}
\item If actor $a$ is an employer, $a \in C$, then determine
the number of workers to fire, $u$, according to the formula:
\begin{equation}
u = max( |W_{a}| - [\frac{m_{a}}{\langle w \rangle}], 0)
\end{equation}
In other words, no workers are fired if the {\it ex ante} wage
bill is payable from the firm's current money holdings
(the wage bill is calculated from the average wage and the
number of employees). Otherwise, the firm's workforce is
reduced to a size such that the wage bill is payable.
\item Select $u$ actors from the set of employees,
$W_{a} = \{ x : e_{x} = a, 1 \leq x \leq N \}$, according
to a uniform distribution. For each selected actor, $c$,
set $e_{c} = 0$ (i.e., fire them).
\end{enumerate}
\end{quotation}
In this model there are no skill differences therefore each actor is
identical. It does not matter which particular workers
are fired, simply the amount, and so the particular individuals to
fire are chosen randomly. Note the asymmetry between hiring
and firing: hiring occurs one individual at a time at a frequency
determined by the number of unemployed actors, whereas firing
may occur in bulk at a frequency determined by the number of firms.
Just as new firms may form when two actors enter an employee-employer
relationship, existing firms may cease trading when all employees
are fired and the capitalist owner enters the unemployed class.

\subsection{Wage payment}

Employers pay wages according to the following rule, which
implements the transfer of value from
capitalist to worker.
\begin{quotation}
\underline{Wage payment rule ${\bf W_{1}}(a)$}: (Stochastic).
\begin{enumerate}
\item For each actor $c$ in $a$'s employee set,
$W_{a} = \{ x : e_{x} = a, 1 \leq x \leq N \}$:
\begin{enumerate}
\item transfer $w$ coins from $a$ to $c$, where $w$ is
selected from the discrete interval $[w_{a},w_{b}]$
according to a uniform distribution. (If employee $a$ has
insufficient funds to pay $w$ then $w$ is selected from
the discrete interval $[0,m_{a}]$ according to a uniform
distribution.)
\end{enumerate}
\end{enumerate}
\end{quotation}
In reality wages are not subject to monthly stochastic
fluctuations. A more concrete model would introduce wage
contracts between employer and employee that fix the
individual employee's wage for the duration of employment.
But in the aggregate, for example in terms of the
total wage bill on average payable by a firm, or wage
and profit shares in national income, the existence
of monthly fluctuations in individual wages is not significant,
and allows a considerable simplification of the model.

\subsection{Historical time}

Finally, the above rules are combined and repeatedly executed
to simulate the functioning of the economy over time. The
following simulation rule orders the possible economic actions:
\begin{quotation}
\underline{Simulation rule ${\bf SR_{1}}$}: Allocate $M/N$ coins
to each of the $N$ actors. Set $e_{i}=0$ for all $1 \leq i \leq N$
(i.e., all actors are initially unemployed).
\begin{enumerate}
\item Execute actor selection rule ${\bf S_{1}}$ to select the
active actor $a$.
\item Execute hiring rule ${\bf H_{1}}(a)$.
\item Execute expenditure rule ${\bf E_{1}}(a)$ that augments
the available market value with new expenditure.
\item If $a$ is associated with a firm, execute market sample rule ${\bf M_{1}}(a)$ that
transfers $m$ coins from the market to the firm owner.
\item Execute firing rule ${\bf F_{1}}(a)$.
\item Execute wage payment rule ${\bf W_{1}}(a)$.
\end{enumerate}
\end{quotation}
The application of rule ${\bf SR_{1}}$ can generate a variety
of events. For example, if the active actor is unemployed it
may get hired by an existing firm, or with lower probability
form a new small firm with another unemployed actor. An
employed active actor will generate a market sample for its
employee, which generates revenue bound by the available
market value, itself a function of the stochastic spending
patterns of other actors. If the active actor is a capitalist
owner of a firm it may decide to fire employees if current
revenues do not cover the expected wage bill. If all employees
are fired then the firm ceases trading. Otherwise, the
wage bill is paid, augmenting the spending power of the
working class, which on the next cycle will affect the
available market value that firms compete for, and so on.

A period of one month is defined as the $N$ applications of
rule ${\bf SR_{1}}$. This means that on average wages are paid once
per month.
\begin{quotation}
\underline{One month rule ${\bf 1M}$}:
\begin{enumerate}
\item Execute rule ${\bf SR_{1}}$.
\item Repeat $N$ times.
\end{enumerate}
\end{quotation}
The rule is executed $N$ times to allow each of the $N$ actors
an opportunity to act. But clearly this does not guarantee that
each actor will in fact act within the month: some actors may
act more than once, others not at all. This introduces a degree
of causal slack that is intended to model the fact that in
real economies events do not occur with strict regularity.
In addition, the repeated random selection of active actors during a
simulated month breaks any symmetries that might be introduced
if actors are selected in a regular order. In reality,
economic actions occur both in order and in parallel and
this causal chaos is modelled by noisy selection.

A period of one year, which is the accounting period, is defined
as 12 applications of the one month rule.
\begin{quotation}
\underline{One year rule ${\bf 1Y}$}:
\begin{enumerate}
\item Execute rule ${\bf 1M}$.
\item Repeat 12 times.
\end{enumerate}
\end{quotation}
The time scale in the model is therefore calibrated to real time
via the empirical fact that on average wages are paid once each month.

The set of nine rules,
$\{{{\bf SR_{1}}, {\bf S_{1}},{\bf H_{1}},{\bf E_{1}},{\bf M_{1}},{\bf F_{1}},
{\bf W_{1}},{\bf 1M},{\bf 1Y}\}}$, and three parameters -- the total
coins in the economy $M$, the total number of actors, $N$, and the fixed
wage interval $\omega$ -- constitute the dynamic, computational model of
the social relations of production, or Social Relations (SR) model for short.

\section{Results}

The rules of the computational model are quite simple yet the dynamic
behaviour of the simulation is rich and complex. A full analysis
of the SR model is beyond the scope of this paper. Instead the aim
is provide a broad overview of the empirical range of the model,
demonstrate that it deserves further analysis, and encourage others
to replicate the computational experiment and begin a full
analytical treatment of its dynamics. Results are presented
that show that the model generates many of the important aggregate distributions
of modern capitalist economies, and therefore implicitly provides
candidate explanations of the causal relationships between them.

The total number of coins, $M$, and the total number of actors, $N$, on
condition that $M >> N$, appear to act as scaling parameters and
do not affect the dynamics, unlike the wage interval parameter.
In all reported results, $N=1000$ and $M=100000$, so that the average wealth
in the economy is 100 coins. On {\em a posteriori} grounds the
wage interval is set to $\omega = [10,90]$; hence, the
minimum wage is 10 coins, the average wage 50 coins, half the mean
wealth in the economy, and the highest possible wage never exceeds
the mean wealth. This results in an almost equal split of national wealth
between the two classes, and generates data in good agreement with
empirical data. A full exploration of the consequences of varying
the wage parameter on the aggregate dynamics is postponed to a sequel.

The results are not sensitive to initial conditions and therefore
the behaviour of a single execution is analysed.
The simulation very rapidly self-organises into a stochastic equilibrium
characterised by stationary distributions. Due to the rapidity of
convergence it makes little difference whether measurements are taken
over the duration of the simulation or after the stationary state has been
reached. Unless stated otherwise the model was allowed to run
for 100 simulated years. The simulation does not settle to a motionless
equilibrium but converges to a dynamic equilibrium of ceaseless motion and change.


\subsection{Class distribution}

\begin{figure}[h!]
\begin{center}
\epsfig{file=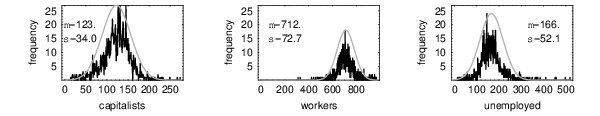,viewport=0 0 280 60,width=1.0\textwidth,clip=true,silent=}
\caption{Class distributions: histograms of the number of actors in each 
economic class with a constant bin size of 1. The smooth lines are fitted 
normal distributions. On average approximately 71.2\% of the population are workers, 
12.3\% are capitalists employing one worker or more, and the remaining 
16.6\% are unemployed.}
\end{center}
\label{fig:classDistribution}
\end{figure}

The social stratification generated by capitalist economies
is a complex phenomenon with systematic causal relations
to the dominant social relations of production. In
reality the social relations of production are more complex
than the relations in the SR model (actors may receive combinations
of wage and property income and therefore belong to more
than one economic class, some actors are self-employed, others
receive the majority of their income from rent, many people
work for governments rather than private enterprises, and so forth).
In consequence, some work is required to map empirical data
on social stratification to the more basic categories employed here.
It is equally clear, however, that the class of capitalists
is numerically small, whereas the class of workers, that is those
actors who predominately rely on wage income for their
subsistence, constitute the vast majority of the population. The
SR model should reflect this empirical fact. The class
breakdown in the SR model can be measured according to the
following rule:
\begin{quotation}
\underline{Class size measure}: After each year
(an application of rule ${\bf 1Y}$) count the number of
workers, $|W|$, capitalists, $|C|$, and unemployed, $|U|$.
\end{quotation}

Figure 1 is a histogram of class sizes generated by the model
collected over the duration of the simulation. The number of workers,
capitalists and unemployed are normally distributed. The normal
distributions summarise a dynamic process that supports social
mobility, where actors move between classes during their imputed
lifetimes, occurring within a stable partition of the population into two
main classes -- a small employing class and a larger employed
class. Fluctuations in class sizes are evidently mean-reverting,
reflecting stable and persistent class sizes, given the exogenous
and constant wage interval. The unemployment rate is higher than
is usually reported in modern economies, but real measures of
unemployment typically discount frictional unemployment,
whereas here all non-employed actors are considered unemployed, and,
in addition, there is no concept of self-employment. In conclusion,
the SR model self-organises into a realistic partition of the working
population into a minority of employers and a majority of employees.

\subsection{Firm size distribution}

\begin{figure}[h!]
\begin{center}
\epsfig{file=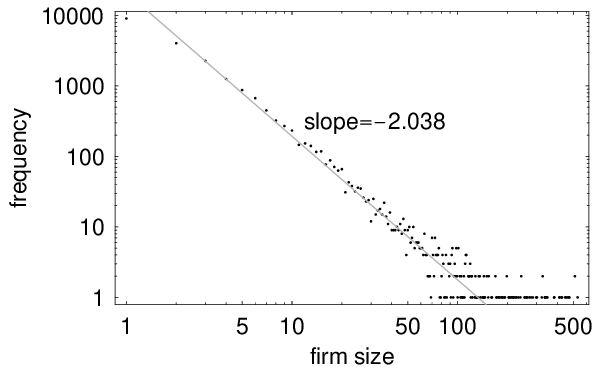,viewport=0 0 300 200,width=0.5\textwidth,clip=true,silent=}
\caption{Firm size distribution: histogram of firm sizes by employees 
in log-log scale with a constant bin size of 1. The straight line is an ordinary least squares regression of the data and represents a power-law distribution
$P(x) \propto x^{-(\alpha+1)}$ with exponent $\alpha=1.038$ for data
collected over 15 simulated years. Axtell \cite{axtell} reports $\alpha=1.059$ 
from data of approximately 5.5 million U.S. firms in 1997. The special 
case $\alpha=1$ is known as the Zipf distribution.}
\end{center}
\end{figure}

Axtell \cite{axtell} analysed US Census Bureau data for US firms trading
between 1988 and 1997 and found that the firm size distribution followed
a special case of a power-law known as Zipf's law, and this relationship
persisted from year to year despite the continual birth and demise of
firms and other major economic changes. During this period the number
of reported firms increased from 4.9 million to 5.5 million.
Gaffeo et. al. \cite{gaffeo03} found that the size distribution of firms
in the G7 group over the period 1987-2000 also followed a power-law,
but only in limited cases was the power-law actually Zipf.
Fuijiwara et. al. \cite{fujiwara03} found that the Zipf law
characterised the size distribution of about 260,000 large firms from
45 European countries during the years 1992--2001. A Zipf law
implies that a majority of small firms coexist with a decreasing
number of disproportionately large firms.

Firm sizes in the SR model are measured according to the following
rule:
\begin{quotation}
\underline{Firm sizes measure}:
After each month (an application of rule {\bf 1M}) count the number
of employees in each firm.
\end{quotation}
The SR model replicates the empirical firm size distribution. Figure
2 is a histogram of firm sizes. The straight line is a fit to the
power-law:
\begin{equation}
P(x) \propto x^{-(\alpha + 1)}
\end{equation}
For data collected over a short time period, such as 15 simulated
years, $\alpha$ approaches $1.0$. The special case
$\alpha=1.0$ is Zipf, and hence the firm size distribution
generated by the model is consistent with the empirical data.
Data collected over shorter periods follows a power-law with exponent
that deviates from 1.

The largest US firm in 1997 had approximately $10^{6}$ employees from a total
reported workforce of about $10^{7}$ individuals \cite{axtell}.
Therefore, the largest firm
size should not exceed about $\frac{1}{10}$th of the total workforce.
Figure 2 shows that, with low but non-zero probability, a single firm
can employ over half the workforce, representing a monopolisation of
a significant proportion of the economy by a single firm, a clearly
unrealistic occurrence. A possible reason for the over-monopolisation
of the economy is the assumption that firms have a single capitalist
owner, which conflates capital concentration with firm ownership.
In reality, large firms normally have multiple owners and individual
capitalists own multiple firms. Further, there are many technical
reasons why particular firms do not grow beyond a certain size that
are ignored in this model. A final point is that the probability of
monopoly within the period of observation decreases with the number
of actors; hence, if the simulation were run with $N=10^{7}$ actors
(which is not possible due to insufficient computational resources)
then it is unlikely that a single firm would employ half the workforce.
Gaffeo et. al. \cite{gaffeo03} note that firms are distributed more
equally during recessions than during expansions, which accounts for
the yearly deviations from Zipf. This relationship has not been tested
in the SR model.

\subsection{Firm growth distributions}

\begin{figure}[h!]
\begin{center}
\epsfig{file=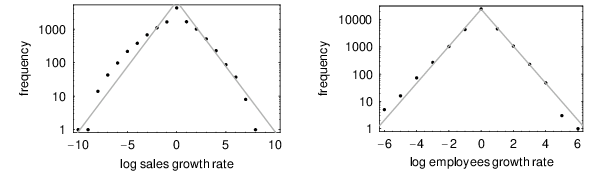,viewport=0 0 300 95,width=1.\textwidth,clip=true,silent=}
\caption{Firm size growth rate distribution: histogram of the log growth rates
of firms per simulated year in linear-log scale with a constant bin size of
1. The LHS graph shows growth rates of firm sales. The RHS graph shows growth
rates of employees. The solid lines are OLS regressions of the data and represent
a Laplace (double-exponential) distribution 
$P(x) \propto e^{-|(x-\alpha)/\beta|}$. Refs. \cite{lee,bottazzi03,stanley96,amaral97b,amaral97,amaral01,fabritiis03} report that log growth rates of sales and employees of US and Italian firms follow a Laplace distribution.}
\end{center}
\label{fig:firmGrowthRateDistribution}
\end{figure}

Stanley et. al. \cite{stanley96} and Amaral et. al. \cite{amaral97b} analyzed
the log growth rates of publicly traded US manufacturing firms in the period
1974 -- 93 and found that growth rates, when aggregated across all sectors,
appear to robustly follow a Laplace (double exponential) form. This holds
true whether growth rates are measured by sales or number of employees.
More precisely, if the annual growth rate is $r=\ln(\frac{s_{t+1}}{s_{t}})$,
where $s_{t}$ is the size of a firm in year $t$, then for all years the
probability density of $r$ is consistent with an exponential decay:
\begin{equation}
f(r) \propto \e^{-|\frac{r-\alpha}{\beta}|}
\end{equation}
with some deviation from the Laplace distribution at high and low growth rates
resulting in slightly `fatter wings' \cite{lee,amaral97,amaral01}.
Bottazzi and Secchi \cite{bottazzi03} replicate these findings and
report a Laplace growth distribution for Italian manufacturing firms
during the period 1989--96.

Firm growth in the SR model is measured according to:
\begin{quotation}
\underline{Firm growth rate measure}:
After each year (an application of rule {\bf 1Y}) calculate
the current size, $s_{t}$, for a firm that traded at some
point during the year. Size is measured in terms of number of employees or total
sales revenue. The growth rate is $s_{t}/s_{t-1}$.
(If a firm ceased trading during the year then $s_{t}=1$,
and if a firm began trading during the year then $s_{t-1}=1$).
\end{quotation}
The SR model generates log annual growth rates for firms
that are consistent with a Laplace distribution, whether growth is
measured in terms of sales or number of employees. Figure 3
plots log growth rates in log-linear scale and reveals the characteristic
`tent' shape signature of a symmetric exponential decay. There is 
some deviation from a Laplace distribution for firms
with shrinking sales, which may be due to noise or represent
some non-accidental property.
Bottazii and Secchi note that standard economic models do not predict a
Laplace growth distribution and propose a stochastic model to explain
the data that, in the abstract, shares similarities to firm growth
in this model. Briefly, both models assume firm growth is a
multiplicative stochastic process constrained by global resources
(this kind of process is embedded in the general dynamics of
the SR model). The SR model is intended to model the social relations
of production, and therefore the replication of the empirical
Laplace growth distribution suggests that the social relations of
production play an important role in constraining the dynamics
of firm growth.

Refs. \cite{lee,fabritiis03,amaral01,amaral97,stanley96} note that
the standard deviation (std) of growth rates decreases as a power law
with size, that is, $\ln \sigma(r) \sim -\beta \ln r$, where
$\beta \approx 0.15$. The SR model does not replicate this finding
given the specified wage interval. In fact, $\ln \sigma(r)$ appears
to increase as a power law with size, although the data is quite
noisy. However, the exponent of the power law is sensitive to
the wage parameter, and it is possible to replicate the
empirical relationship at lower average wages. Explanations
of the relationship between growth variation and size assume
that firms have internal structures such that increased size
lessens market risk \cite{amaral01,amaral97}, which contrasts
with the simple firm structure employed in this model.
Axtell in ref. \cite{axtell99} presents an agent-based model of
the life-cycle of firms that replicates the Zipf size distribution, 
Laplace growth rates and power-law scaling of the std of growth. Firms in Axtell's
model have more complex internal relations between employees
who make optimising decisions on the trade-off between effort
and reward. 

\subsection{Firm demise distribution}
\begin{figure}[h!]
\begin{center}
\epsfig{file=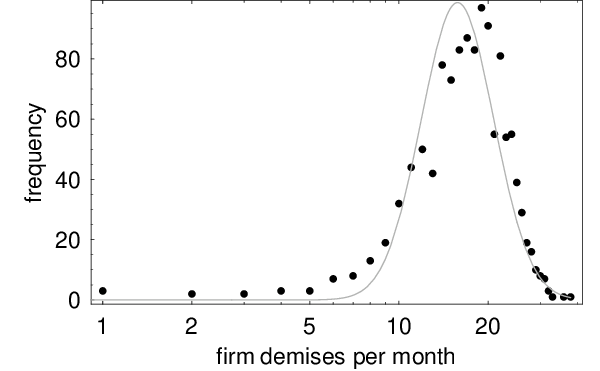,viewport=0 0 300 200,width=0.5\textwidth,clip=true,silent=}
\caption{Firm demises distribution: Histogram of firm demises per simulated month in log-linear scale with a 
constant bin size of 1. The solid line is a fit to the lognormal distribution. Cook and Ormerod \cite{cook03} 
report that the distribution of US firm demises per year during the period 1989 to 1997 is closely 
approximated by a lognormal distribution.}
\end{center}
\label{fig:firmDemiseDistribution}
\end{figure}

Cook and Ormerod \cite{cook03} report that the distribution of
US firm demises per year during the period 1989 to 1997 is closely
approximated by a lognormal distribution, and note that the number of
demises varies little from year to year with no clear connection to
recession or growth. 

The number of firm demises
per month in the simulation is measured according to:
\begin{quotation}
\underline{Firm demise measure}:
After each month (an application of rule {\bf 1M}) count the number
of firm demises that occurred during the month. A firm demise occurs when a firm fires
all its employees.
\end{quotation}
Demises per month are measured rather than per year in order to
avoid bucketing the data. Figure 4 is a histogram of firm demises
per month with a fitted lognormal distribution. It shows that
the model generates a distribution of firm demises that is
approximated by a lognormal distribution
and is therefore consistent with empirical findings.

According to Cook and Ormerod the average number of firms in the US during
the period 1989 to 1997 was 5.73 million, of which on average
611,000 died each year. So roughly 10\% of firms die each year.
In the simulation on average 18 firms die each month
and therefore on average 216 firms die each year, a figure
in excess of the 123 firms that exist on average. So although
the distribution of firm demises is consistent with empirical
data, the rate at which firms are born and die is much higher
than in reality. This is not too surprising when it is considered
that the model entirely abstracts from the material nature of
the goods and services processed by the economy and any
persistent demand for them. In this model firms compete
by playing a game of chance that models the unpredictability
of a competitive economy. But the complete absence of
the material side of the economy results in an unrealistic
level of volatility in market interactions. The SR model
must therefore be extended to include causal relations between
the social architecture and the forces of production.
Clearly there is a limit to what may be deduced from consideration
of the social relations of production alone.

\subsection{GDP growth}
\begin{figure}[h!]
\begin{center}
\epsfig{file=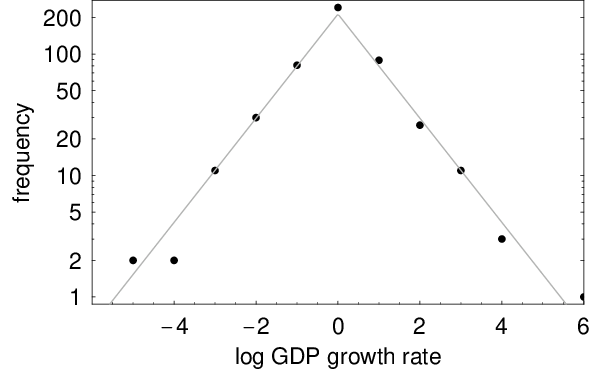,viewport=0 0 300 200,width=0.5\textwidth,clip=true,silent=}
\caption{Rescaled GDP growth rate distribution: histogram of the log growth rate
of GDP in linear-log scale with a constant bin size of 1. The solid lines are 
OLS regressions of the data and represent a Laplace (double-exponential) distribution 
$P(x) \propto e^{-|(x-\alpha)/\beta|}$. Ref. \cite{lee} reports that log GDP growth
rates of 152 countries during the period 1950--92 follow a Laplace distribution.}
\end{center}
\label{fig:GDPGrowthRateDistribution}
\end{figure}

Gross Domestic Product (GDP) measures the value of gross production
at current prices, including consumption and gross investment.
Lee at. al. \cite{lee} and Canning et. al. \cite{canning98}
analyse the GDP of 152 countries during the period 1950--52
and find that the distribution of GDP log growth rates is
consistent with a Laplace distribution, and therefore conclude that
firm growth and GDP growth are subject to the same laws \cite{lee}.

The GDP growth rate in the SR model is measured according to the
following rule:
\begin{quotation}
\underline{GDP growth measure}: At the close of year $t$ calculate
the total firm income, $X_{t}$, received during that year (i.e.,
all income received during application of rule ${\bf M_{1}}$).
The GDP growth rate is then $X_{t}/X_{t-1}$.
\end{quotation}
Empirical measurements of GDP must be detrended to remove the
effects of inflation but this is unnecessary when measuring GDP
in the model due to the assumption of a fixed amount of money.

Figure 5 plots log GDP growth rate for the simulated economy in log-linear scale.
The data is noisy but consistent with a Laplace distribution when sampled over
a period of 100 years so for clarity figure 5 contains data from an extended run
of 500 years. The characteristic tent shape indicates that the SR model is consistent
with the Laplace distribution of GDP growth. 

Gatti et. al. in ref. \cite{gatti03}
present an agent-based model of the life-cycle of firms that replicates the Zipf 
size distribution and Laplace growth rates of firms and
aggregate output (GDP). They show that the power-law of firm size
implies that growth is Laplace distributed and also that small
micro-shocks can aggregate into macro-shocks to generate recessions.
Firms in their model are not disaggregated into employees and
employers and market shocks are exogenous, whereas in this model 
firms are composed of individuals and are subject to 
endogenous shocks that are the consequence of the competition 
for a finite amount of available market value, itself a product of income flows.

\subsection{Duration of recessions}
\begin{figure}[h!]
\begin{center}
\epsfig{file=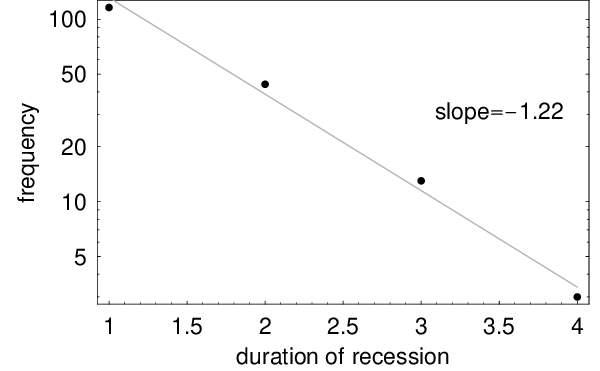,viewport=0 0 300 200,width=0.5\textwidth,clip=true,silent=}
\caption{Recession duration distribution: Histogram of the frequency of the
duration of recessions in log-linear scale with a constant bin size of 1.
The solid line is a fit to an exponential distribution, 
$f(d) \propto \lambda e^{-\lambda d}$, with exponent
$\lambda = 1.22$, representing an average recession duration of 1.22 
simulated years.}
\end{center}
\label{fig:GDPGrowthRateDistribution}
\end{figure}

Wright \cite{wright03b}, reinterpreting empirical data presented 
by Ormerod and Mounfield \cite{ormerod01}, concludes that the frequency of the
duration of economic recessions, where a recession is defined as a
period of shrinking GDP, follows an exponential law for 17 Western
economies over the period 1871--1994. Recessions tend not to last
longer than 6 years, the majority of recessions last 1 year,
and for the US the longest recession has been only 4 years \cite{ormerod02}.

The duration of recessions in the simulation is measured according
to the following rule:
\begin{quotation}
\underline{Recession duration measure}: A recession begins
when
\begin{quotation}
$X_{t}/X_{t-1} < 1$
\end{quotation}
 and ends when
\begin{quotation}
$X_{t+k}/X_{t+k-1} \geq 1$
\end{quotation}
The duration of the recession is $k$ years.
\end{quotation}
The SR model is in close agreement with these empirical findings.
Figure 6 is a histogram of the frequency of the duration of
recessions collected over a period of 500 simulated years.
The functional form of the frequency of duration of recessions
is exponential, $f(d) \propto \lambda \e^{-\lambda d}$,
with $\lambda=1.22$, which compares to a value of $\lambda=0.94$
for the empirical data \cite{wright03b}. The value of $\lambda$ is
the average duration of a recession. Also, the duration of
recessions in the model ranges from 1 to 4 simulated years.

\subsection{Income shares}

GDP is the sum of revenues received by firms
during a single year. Firms pay the total wage bill, $W$, from this revenue.
Hence the total value of domestic output is divided into a share that workers receive
as wages, $X_{w}=\frac{W}{X}$, and the remainder that capitalists
receive as profit, $X_{p} = 1 - X_{w}$. Advanced capitalist countries
publish national income accounts that allow wage and profit shares
to be calculated, which reveal some characteristic features.
Shares in national income have remained fairly stable during
the twentieth century, despite undergoing yearly fluctuations.
For example, the profit share, normally lower than the wage share,
is between 0.25 to 0.4 of GDP, although it occasionally can be
as high as 0.5 (source: Foley and Michl's calculations in 
ref. \cite{foley99} for US, UK and Japan spanning a period of 
over 100 years; other authors place the wage share nearer to $\frac{1}{2}$, for example on
average 0.54 between 1929 and 1941 for the USA \cite{kalecki54}
and similar in chapters 3 and 8 of ref. \cite{farjoun}).

\begin{figure}[t!]
\begin{center}
\epsfig{file=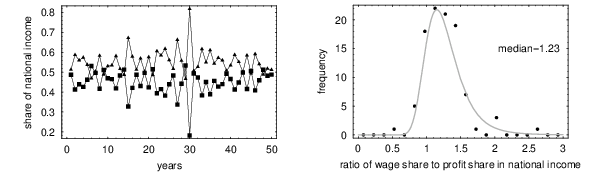,viewport=0 0 300 100,width=1.0\textwidth,clip=true,silent=}
\caption{Wage and profit shares in national income. The LHS graph is a representative
time series of the fluctuating shares in national income. GDP, denoted $X$, 
is the sum of revenues received by firms during a single year. The solid triangles
are the wage share, $X_{w}$, which represents the sum total of wages paid
to the working class, $W$, divided by GDP, $X_{w}=\frac{W}{X}$. The 
solid squares are the profit share, $X_{p}$, which represents the sum total 
of profits received by the capitalist class divided by GDP, 
$X_{p}=1-\frac{W}{X}$. The wage share fluctuates around a mean of
0.55 and the profit share fluctuates around a mean of 0.45. The RHS graph 
is a histogram of the ratio $\frac{W}{1-W}$. The smooth line is a fitted 
pdf of the ratio of two normal variates, which indicates that 
fluctuations of shares in national income are normally distributed 
around long-term stable means.}
\end{center}
\label{fig:moneyHoldings}
\end{figure}

Income shares in the simulation are measured according to:
\begin{quotation}
\underline{Income shares measure}:
Calculate the GDP, $X$, for the year.
Count the total income received by workers during the
same year (i.e.,
income received during application of rule ${\bf W_{1}}$), which
is defined as the total wage bill $W$.
The wage share is $X_{w}=\frac{W}{X}$, and the profit share
is $X_{p} = 1 - X_{w}$.
\end{quotation}
Figure 7 is a plot of the shares in national income generated by the
model. The profit share is generally lower than the wage share,
and the yearly fluctuations are normally distributed about
long-term stable values. Ignoring differences of
definition, and for the purposes of a rough and ready comparison,
the model generates an average profit share of 0.45, which compares
well to the empirical data. The model therefore reproduces the
empirical situation of fluctuations about a long-term stable
mean, and additionally the profit and wage shares have realistic
values, although it is an open question whether suitably
de-trended fluctuations are normally distributed in capitalist
economies.

\subsection{Disaggregated income distributions}

Income shares can be disaggregated and measured at the level
of individuals in order to understand income differentiation
within classes. The empirical income distribution is characterised
by a highly unequal distribution of income, in which a very small
number of households receive a disproportionate amount of
the total (e.g., using wealth as an indicator of income,
in 1996 the top 1\% of individuals in the US owned 40\% of the total
wealth \cite{levy97}). The higher, property-income, regime of the
income distribution can be fitted to a Pareto (or power) distribution
\cite{levy-wealth,levy97,matteo03,dragulescu02b,nirei03,nirei03b,souma00},
whereas the lower, or wage-income, regime, which
represents the vast majority of the population, is normally fitted
to a lognormal distribution \cite{souma00,montroll83,badger80}, but recently
some researchers \cite{nirei03b,dragulescu02b,dragulescu02a}
report that an exponential (Boltzmann-Gibbs) distribution
better describes the empirical data.
Plotting the income distribution as a complementary cumulative
distribution function (ccdf) in log-log scale reveals a
characteristic `knee' shape at the transition between the
two regimes \cite{matteo03,dragulescu02b,dragulescu02a,souma00,nirei03b}.

\begin{figure}
\centering
\subfigure[The complete income distribution plotted as a ccdf in log-log scale. The data is binned at a constant size of 1. Note the characteristic `knee' shape, a feature found in empirical distributions. The transition from the lognormal to the Pareto regime occurs between $P(x)=0.1$ and $P(x)=0.01$, which means that under 10\% of incomes follow the Pareto law.]{\epsfig{file=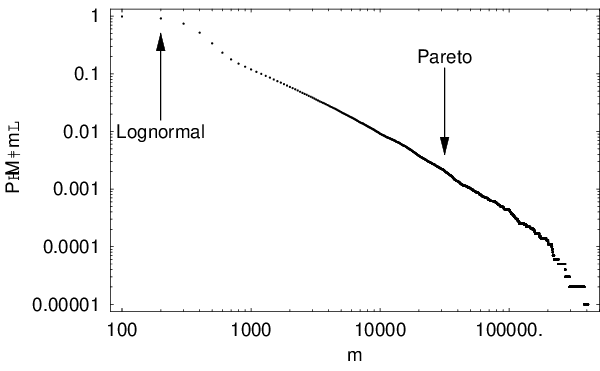,viewport=0 0 300 200,width=0.47\textwidth,clip=true,silent=}}\qquad
\subfigure[The class components of the income distribution plotted as ccdfs in log-log scale. Note the long tail of the capitalist income distribution. Worker income is clustered around the average wage.]{\epsfig{file=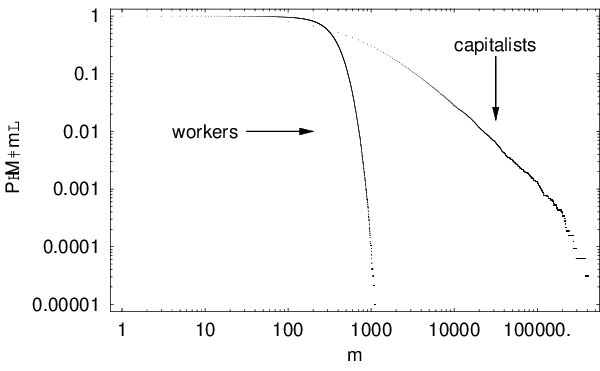,viewport=0 0 300 200,width=0.47\textwidth,clip=true,silent=}}\\
\subfigure[The lower regime of the income distribution plotted in log-linear scale. The solid line is a fit to a lognormal distribution. The approximately lognormal distribution results from a mixture of wage income and small employer income.]{\epsfig{file=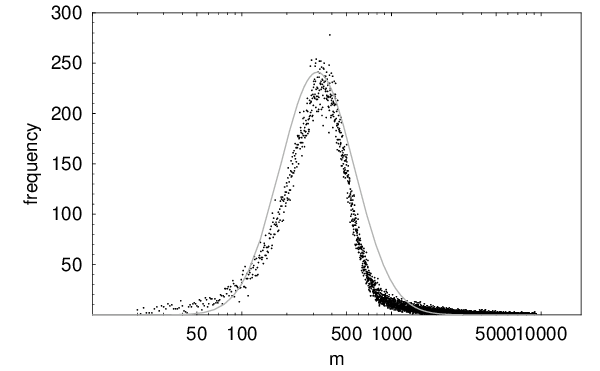,viewport=0 0 300 200,width=0.47\textwidth,clip=true,silent=}}\qquad
\subfigure[The power law regime of the income distribution plotted as a ccdf in log-log scale. The straight line is a fit to the power (Pareto) law, $P(x) \propto x^{-(\alpha + 1 )}$, where $\alpha = 1.3$.]{\label{a label}\epsfig{file=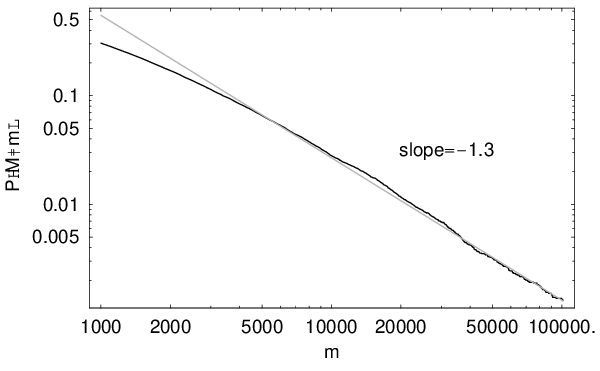,viewport=0 0 300 200,width=0.47\textwidth,clip=true,silent=}}%
\caption{Graphical analysis of the stationary income distribution. Incomes are measured over the duration of a year and the data is collected over the duration of the simulation.}
\label{incomeDistribution}
\end{figure}

The functional form of the income distribution is stable over many years,
although the parameters seem to fluctuate within narrow bounds.
For example, for property-income, the power-law, $P(x) \propto x^{-(\alpha+1)}$,
has a value $\alpha=1.3$ for the UK in 1970 \cite{levy-wealth},
$\alpha=[1.1,1.3]$ for Australia between 1993 and 1997 \cite{matteo03},
$\alpha=1.7$ for US in 1998 \cite{dragulescu02b}, on average $\alpha=1.0$
for post-war Japan \cite{nirei03}, and $\alpha=[0.5,1.5]$ for US and
Japan between 1960 and 1999 \cite{nirei03b}. In sum, the income distribution
is asymptotically a power-law with shape parameter $\alpha \approx 1.0$,
and this regime normally characterises the top 1\% to 5\% of incomes.

The two-parameter lognormal distribution
\begin{equation}
P(x) = \frac{1}{x \sigma \sqrt{2 \pi}} \exp(\frac{-(\log \frac{x}{m})^{2}}{2 \sigma^{2}})
\end{equation}
where $m$ is the median, and $\beta = 1/\sqrt{2 \sigma^{2}}$ is
the Gibrat index, can describe the remaining 95\% or so of incomes.
For example, for post-war Japan, the Gibrat index ranges between approximately
$\beta=2.25$ and $\beta=3.0$ \cite{souma00}. In contrast, if the lower income
range is fitted to an exponential law
\begin{equation}
P(x) \propto \lambda \e^{\lambda x}
\end{equation}
then by analogy with a perfect gas, from which the Boltzmann-Gibbs
law originates, $\lambda$ is interpreted as an average
economic `temperature', which should be close to the average wealth
in the economy, adjusting for the effects of the Pareto tail.

The SR model is in close qualitative and quantitative agreement with
all these empirical facts. It also explains why there are two
major income regimes, and provides a candidate explanation of
why the distribution of low incomes is sometimes identified as
either lognormal or exponential.

Incomes in the simulation are measured according to the following rule:
\begin{quotation}
\underline{Incomes measure}: After each year (an application of
rule ${\bf 1Y}$) calculate the total income received by each
actor during the year. Both wage income (from rule ${\bf W_{1}}$) and
capitalist income (from rule ${\bf M_{1}}$) are counted as income.
\end{quotation}
Figure 8(a) is a plot of the stationary income ccdf
generated by the model. It reproduces the characteristic `knee' shape found
in empirical income distributions. The `knee' is
formed by the transition from the lower regime,
consisting mainly of the wealth of the working class
and owners of small firms, to the higher regime, consisting
mainly of the wealth of the capitalist class. The knee occurs
at around $P(M \geq m)=0.1$, which means the power-law
regime holds for at most 10\% of incomes.

Figure 8(b) splits the income distribution according to class.
The capitalist distribution has a long tail, qualitatively
different from the worker distribution, which is clustered
around the average wage. Figure 8(c) is a plot of the lower
regime of the income distribution in log-linear scale
fitted to a lognormal distribution with Gibrat
index $\beta=1.42$. Figure 8(d) is a plot of the
property-income regime in log-log scale. The straight line
fit indicates that higher incomes asymptotically approach
a power-law distribution of the form $P(x) \propto x^{-(\alpha+1)}$,
with $\alpha=1.3$.
The two income regimes are consequences of the two major sources
of income in capitalist societies, that is wages
and profits, and the overall income distribution is a
mixture of two qualitatively different distributions.
The lower regime is fitted better by a lognormal
distribution rather than an exponential. The lognormal
distribution, in this model, is not the result of stochastic
multiplicative process, which is the explanation often proposed,
but results from a mixture of normally distributed
wage incomes and the profit-income of small firm
owners. It is an open question whether the lognormal distribution
found in empirical data can be similarly explained by
the combined effect of income from employment and the income of
small employers.

At first glance it appears that the model contradicts
empirical evidence that the lower income regime is exponentially
distributed. But if the stationary distribution of
money holdings (i.e., instantaneous wealth) is measured, rather than income,
a different picture emerges, which may help explain the
lack of consensus in empirical studies. Wealth is measured according
to the following rule:
\begin{quotation}
\underline{Wealth measure}: After each year (an application of
rule ${\bf 1Y}$) calculate the total money held by each
actor.
\end{quotation}
Figure 9 is a plot of the
stationary money ccdf generated by the model. As before,
figure 9(a) reproduces the characteristic `knee' shape found
in empirical income distributions. But in this case the lower
regime is characterised by an exponential (or Boltzmann-Gibbs)
distribution. The transition between regimes occurs approximately
in the middle of the ccdf corresponding to a situation in which
the total wealth in the economy is distributed approximately
evenly between the classes. Figure 9(c) plots the workers'
money distribution in log-linear scale. The straight line fit
reveals an exponential distribution of the form
$P(x)=\lambda \e^{\lambda x}$, where $\lambda = 0.017$, which
is indeed close to the average wealth in the economy,
$\lambda = \frac{M}{N}=0.01$ \cite{dragulescu}.
Figure 9(d) plots the capitalists'
money distribution in log-log scale. The straight line fit reveals a
power-law distribution with similar exponent to that of income.

The higher income and wealth regimes are qualitatively identical,
but the lower income and wealth regimes are qualitatively
distinct. Measuring the lower end of income yields a lognormal 
distribution, whereas measuring the lower end of wealth yields an 
exponential. Income depends solely on monies received during an 
accounting period, whereas wealth depends on both income and 
spending patterns. The differences between the empirical studies 
could be due to differences in whether the measures employed are 
predominately income measures or wealth measures.

\begin{figure}
\centering
\subfigure[The complete money distribution plotted as a ccdf in log-log scale. The transition from the Boltzmann-Gibbs to Pareto regime occurs in the middle of the ccdf. The data is binned at a constant size of 1.]{\epsfig{file=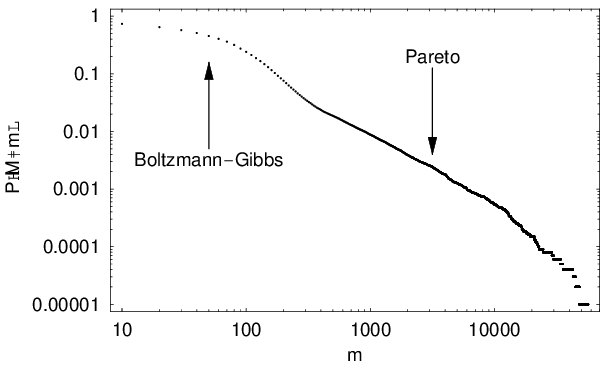,viewport=0 0 300 200,width=0.47\textwidth,clip=true,silent=}}\qquad
\subfigure[The class components of the money distribution plotted as ccdfs in log-log scale. Note the long tail of the capitalist money distribution.]{\epsfig{file=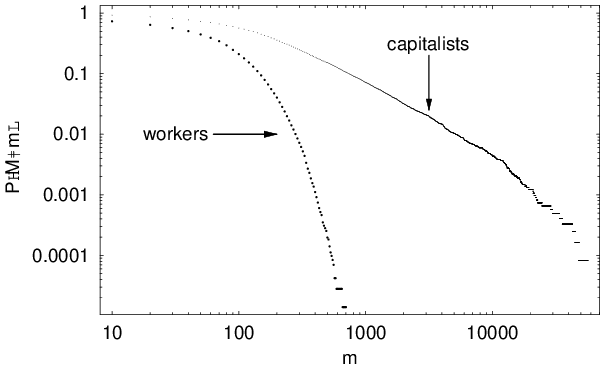,viewport=0 0 300 200,width=0.47\textwidth,clip=true,silent=}}\\
\subfigure[A section of the workers' money ccdf plotted in linear-log scale. The straight line is a fit to the exponential (Boltzmann-Gibbs) law, $P(x)=\lambda e^{\lambda x}$, where $\lambda=0.017$.]{\epsfig{file=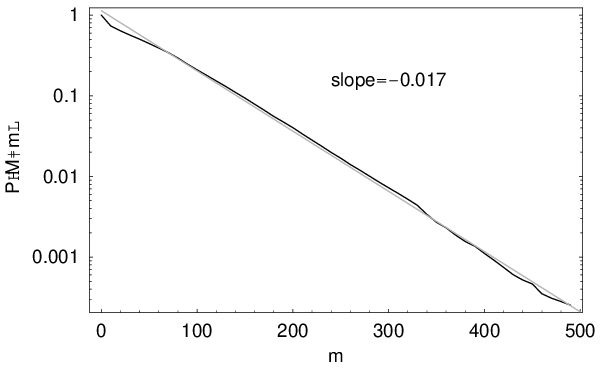,viewport=0 0 300 200,width=0.47\textwidth,clip=true,silent=}}\qquad
\subfigure[A section of the capitalists' money ccdf plotted in log-log scale. The straight line is a fit to the power (Pareto) law, $P(x) \propto x^{-(\alpha+1)}$, where $\alpha=1.3$.]{\label{a label}\epsfig{file=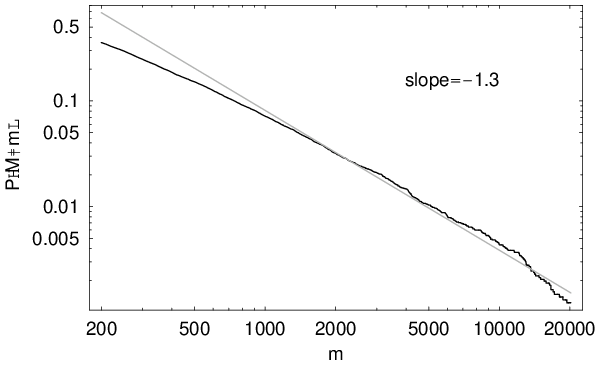,viewport=0 0 300 200,width=0.47\textwidth,clip=true,silent=}}%
\caption{Graphical analysis of the stationary money distribution. Money holdings are measured at the end of a year and the data is collected over the duration of the simulation.}
\label{moneyDistribution}
\end{figure}

The lognormal and power-law fits are only approximations to
the true distributions, and a full analysis of the income
distribution is postponed. However, a few brief points
can be made. A popular explanation of the power-law tail of the
income distribution is that it arises from an underlying stochastic multiplicative
process, often thought to model the geometric growth of capital
invested in financial markets
\cite{nirei03b,nirei03,reed00,reed00b,levy-wealth,levy97,bouchaud00}.
The importance of financial markets in determining
capital flows and hence capitalist income is undeniable.
But the model developed here shows that an income power-law
can arise from industrial capital invested in firms, absent
financial markets that support capital reallocation between
industries or between capitalists. Capitalist income, in
this model, is not derived from investment in portfolios
that provide a return, but is composed of the sum of values added via
the employment of productive workers. In this sense, capitalist income
is additive, not multiplicative. But workers are grouped in
firms that follow a power-law of size. Hence, the power-law of capitalist income
is due to a power-law in the network structure of the wage-capital relation.
Di Matteo et. al. \cite{matteo03} show that an additive stochastic
model of interacting agents on a power-law network generates power-law
distributions, and presumably a similar explanation accounts for
the capitalist income distribution generated by this model.
Clearly, the firm ownership structure in this model is highly simplified,
and it is an open question whether this conclusion holds in models
that include joint ownership of firms.

\subsection{Rate-of-profit distribution}
\begin{figure}[h!]
\begin{center}
\epsfig{file=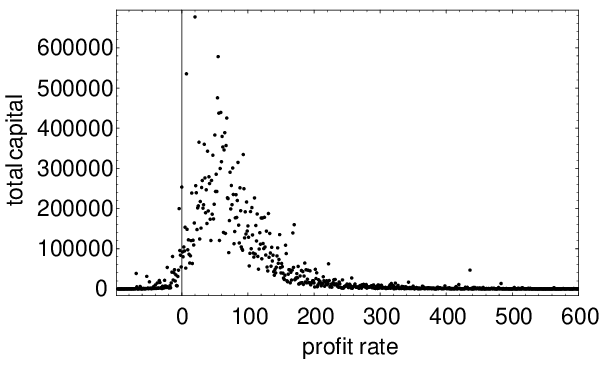,viewport=0 0 300 200,width=0.5\textwidth,clip=true,silent=}
\caption{Capital-weighted rate-of-profit distribution: Histogram of amount of capital invested that generated a given percentage profit rate within a simulated year. The data is collected over the duration of the simulation and binned at a constant size of 1. The average profit rate is 80.5\% and the median profit rate is 64\% (on average 1 coin invested returns 1.8 coins). Wells \cite{wells01} measured the profit rate distribution of over 100,000 UK firms trading in 1981 and found that the distribution was right-skewed.}
\end{center}
\label{fig:profitRateDistribution}
\end{figure}

Farjoun and Machover \cite{farjoun} propose that the proportion of
industrial capital (out of the total capital invested in the economy)
that finds itself in any given rate-of-profit bracket will be
approximated by a gamma distribution by analogy with the distribution
of kinetic energy in a gas at equilibrium. The gamma distribution is
a right-skewed distribution. Wells \cite{wells01} measured the
rate-of-profit distribution of over 100,000 UK firms trading in
1981 and found that in general the distribution was right-skewed, whatever precise
definition of profit was used, although the data was not well
characterised by a gamma distribution.

In reality capitalist owners of firms invest in both variable
(wages) and constant capital (investment in commodity inputs to
the production process and relatively long-lasting means of production)
\cite{okishio87} and the rate-of-profit is calculated on the total
capital invested. The SR model abstracts from the forces
of production and hence capitalist owners invest only in variable
capital (i.e. expenditures on wages due to the application of
rule ${\bf W_{1}}$). Capitalists also spend income in
the marketplace according to rule ${\bf E_{1}}$, and this
expenditure could be interpreted as either consumption or investment
in constant capital, but to theoretically ground the latter interpretation
the model would need to be extended to include a determination of
the distribution of ratios of constant to variable capital across
firms. Rather than introduce the material side of the economy,
which properly belongs to future substantive extensions of the model,
the rate-of-profit in the simulation is calculated on variable
capital alone. Hence rate-of-profit measures will exceed those
found empirically.

The rate-of-profit distribution in the model
is measured according to:
\begin{quotation}
\underline{Profit rate measure}: After each year (an
application of rule ${\bf 1Y}$) calculate the profit
rate for each firm trading at the close of the year.
The profit rate, $p_{i}$, of firm $i$ is defined as
\begin{equation}
p_{i} = 100 \left( \frac{r_{i}}{w_{i}} - 1 \right)
\end{equation}
where $r_{i}$ is the total revenue received during the year
and $w_{i}$ is the total wages paid during the year.
\end{quotation}

\begin{figure}[t!]
\begin{center}
\epsfig{file=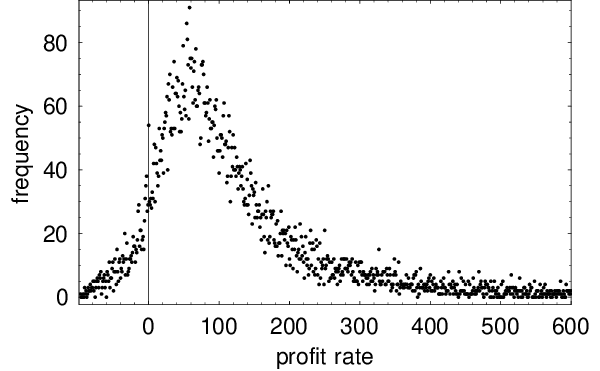,viewport=0 0 300 200,width=0.5\textwidth,clip=true,silent=}
\caption{Firm-weighted rate-of-profit distribution: Histogram of number of firms that generated a given percentage profit rate within a simulated year. The data is collected over the duration of the simulation and binned at a constant size of 1. Wells \cite{wells01} measured the firm-weighted profit rate distribution of over 100,000 UK firms trading in 1981 and found that, similar to the capital-weighted rate-of-profit, the distribution was right-skewed, although less noisy.}
\end{center}
\label{fig:profitRateDistribution}
\end{figure}

Figure 10 graphs the amount of capital that returned a
given profit within a year. Consistent with empirical research
the distribution is highly right-skewed. Wells \cite{wells01}
reports that if the rate-of-profit is weighted according to
number of firms, rather than capital invested, the distribution
is also right-skewed and very similar in overall character, 
although less noisy. Figure 11 graphs the firm-weighted distribution
from the simulation. It is also right-skewed, like the 
capital-weighted distribution, but considerably less noisy. 
The SR model provides the opportunity to deduce
an analytical form for the industrial rate-of-profit 
distribution given additional assumptions on capital investment.
As a step toward this goal an approximation to the firm-weighted 
rate-of-profit distribution in the SR model is now derived.

Consider a single firm that trades for a single year and 
has an average size of $s$ employees during
this period. The firm samples the market on average $12 s$
times during a year. This is a simplification, as during a year, firms
are created and destroyed, and therefore do not necessarily
interact with the marketplace over the whole year.
The value of each market sample, $M_{i}$,
is a function of the instantaneous money distribution, which is mixture 
of exponential and Pareto forms. Assume each $M_{i}$ is independent
and identically distributed (iid) with mean $\mu_{m}$ and variance $\sigma_{m}^{2}$.
During a month the same employee may be repeatedly selected,
or not selected, due to the causal slack introduced by rule ${\bf 1M}$.
Therefore the value generated per employee per month, $V_{i}$,
is some function $f$ of $M_{i}$ independent of the firm size $s$. 
Simplifying further to avoid detailed consideration of the distribution
of market samples per employee, assume that $f(x) = x + v$, where $v$ is a
constant. Hence each $V_{i}$ is idd with mean $\mu_{1} = \mu_{m} + v$
and variance $\sigma_{1}^{2} = \sigma_{m}^{2}$.
By the Central Limit Theorem the sum of the firm's market samples in 
a year, which constitutes the total revenue, $R$, can be approximated by 
a normal distribution 
$R = \sum V_{i} \approx N( \mu_{r}, \sigma_{r}^{2} )$,
where $\mu_{r} = 12 s \mu_{1}$ and $\sigma_{r}^{2} = 12 s \sigma_{1}^{2}$.

The firm's total wage bill for the year, $W$, is the sum of 
$12 (s - 1)$ individual wage payments, $\omega_{i}$.
Note that the capitalist owner does not receive wages. Each
$\omega_{i}$ is iid according to a uniform distribution, 
$\omega_{i} \sim U( \omega_{a}, \omega_{b})$, with mean
$\mu_{2} = (\omega_{a} + \omega_{b})/2$ and variance 
$\sigma_{2}^{2} = (\omega_{b} - \omega_{a})^{2}/12$. 
By the Central Limit Theorem the wage bill, $W$, can be approximated
by a normal distribution
$W = \sum \omega_{i} \approx N( \mu_{w}, \sigma_{w}^{2})$,
where $\mu_{w} = 12 ( s - 1) \mu_{2}$ and 
$\sigma_{w}^{2} = 12 (s - 1) \sigma_{2}^{2}$.

Define the ratio of revenue to the wage bill as $X=R/W$ and assume that
$R$ and $W$ are independent. $X$ is the ratio of two normal variates
and its pdf is derived by the transformation method to give:
\begin{eqnarray}
\nonumber f_{X}(x \mid s) & = & \frac{\exp \left[ -\frac{1}{2}(\frac{\mu_{r}^{2}}{\sigma_{r}^{2}} + \frac{\mu_{w}^{2}}{\sigma_{w}^{2}}) \right] }{4 \pi k_{1}^{3/2} } \\
& & \left( 2 \sigma_{r} \sigma_{w} \sqrt{k_{1}} + 
\e^{\Lambda(x)} \sqrt{2 \pi} (1 + \Phi(\sqrt{\Lambda(x)})) 
(\mu_{w} \sigma_{r}^{2} + x \mu_{r} \sigma_{w}^{2})
\right)
\end{eqnarray}
where
\begin{eqnarray}
\nonumber k_{1} & = & \sigma_{r}^{2} + x^{2} \sigma_{w}^{2} \\
\nonumber \Lambda(x) & = & \frac{(\mu_{w} \sigma_{r}^{2} + x \mu_{r} \sigma_{w}^{2})^{2}}
{2 \sigma_{r}^{2} \sigma_{w}^{2} (\sigma_{r}^{2} + x^{2} \sigma_{w}^{2})} \\
\nonumber \Phi(x) & = & \frac{2}{\sqrt{\pi}} \int_{0}^{x} \e^{-t^{2}} \d t
\end{eqnarray}

(8) is the pdf of the rate-of-profit conditional on the firm size $s$. 
The unconditional rate-of-profit distribution can be obtained by 
considering that firm sizes are distributed according to a Pareto 
(power-law) distribution:
\begin{eqnarray}
\nonumber f_{S}(s) = \frac{\alpha \beta^{\alpha}}{s^{\alpha + 1}}
\end{eqnarray}
where $\alpha$ is the shape and $\beta$ the location parameter. Firm sizes
in the model range between 1 (a degenerate case of an unemployed
worker) to a maximum possible size $N$. Therefore the truncated
Pareto distribution
\begin{eqnarray}
\nonumber g_{S}(s) = f_{S}( s \mid 1 < S \leq N ) = \frac{f(s)}{F(N) - F(1)} = 
\frac{s^{-(1 + \alpha)} \alpha}{1 - N^{-\alpha}}
\end{eqnarray}
where 
\begin{eqnarray}
\nonumber f(s) = F'(s)
\end{eqnarray}
is formed to ensure that all the probability mass is between 1 and $N$.
By the Theorem of Total Probability the unconditional distribution
$f_{X}(x)$ is given by:
\begin{equation}
f_{X}(x) = \int_{2}^{N} f_{X}(x \mid s) g_{S}(s) \d s
\end{equation}
where the range of integration is between 2 and $N$ as firms of size
1 are a degenerate case that do not report profits. Expression
(9) defines the $g_{S}(s)$ parameter-mix of $f_{X}(x \mid S = s)$.
The rate-of-profit variate is therefore composed of a parameter-mix
of a ratio of independent normal variates each conditional on a 
firm size $s$ that is distributed 
according to a power-law. Writing (9) in full yields:
\begin{eqnarray}
\nonumber f_{X}(x) & = & \int_{2}^{N}               
\frac{\exp \left[ -6(\frac{s \mu_{1}^{2}}{\sigma_{1}^{2}} + \frac{(s-1) \mu_{2}^{2}}{\sigma_{2}^{2}}) \right] }{2 \pi \Theta^{3/2}(x) } \\
\nonumber & &  \left( k_{2} \sqrt{\Theta(x)} + \sqrt{6 \pi} \Psi(x) \exp \left[ \frac{6 \Psi^{2}(x)}{k_{2}^{2} \Theta(x)} \right] \left( 1 + \Phi(\frac{ \sqrt{6} \Psi(x)}{k_{2} \sqrt{\Theta(x)}} ) \right) 
\right) \\
& & \frac{s^{-(1 + \alpha)} \alpha}{1 - N^{-\alpha}} \; \d s
\end{eqnarray}
where
\begin{eqnarray}
\nonumber k_{2} & = & \sqrt{s \sigma_{1}^{2}} \sqrt{(s-1) \sigma_{2}^{2}} \\
\nonumber \Theta(x) & = &  s \sigma_{1}^{2} + (s - 1) x^{2} \sigma_{2}^{2} \\
\nonumber \Psi(x) & = & (s - 1) s (\mu_{2} \sigma_{1}^{2} + x \mu_{1} \sigma_{2}^{2} )
\end{eqnarray}
(10) is the pdf of $X=R/W$ but the rate-of-profit in the simulation
is measured as $P = 100 \left( X - 1 \right)$. The pdf of P is therefore a 
linear transform of X:
\begin{equation}
f_{P}(x) = \frac{1}{100} f_{X}( 1 + \frac{x}{100} )
\end{equation}
(11) defines a distribution with 6 parameters:
(i) the mean employee market sample $\mu_{1}$, (ii) the variance of the employee
market sample $\sigma_{1}^{2}$, (iii) the mean wage $\mu_{2}$, (iv) 
the wage variance $\sigma_{2}$, (v) the Pareto exponent, $\alpha$, 
of the firm size distribution, and (vi) the number of economic actors
$N$.

(11) is solved numerically to compare the distribution of the 
theoretically derived variate $P$ with the profit data generated by the SR model.
The values of the parameters are measured from the simulation. In this
particular case $\mu_{1} = \mu_{m} + v \approx 50 + v$, 
$\sigma_{1}^{2} = \sigma_{m}^{2} \approx 55000$,
$\mu_{2} \approx 50$, $\sigma_{2}^{2} \approx 533.3$, $\alpha \approx 1.04$
and $N = 1000$. The best fit is achieved with $v = 25$ coins. Figure 12 plots 
the pdf $f_{P}(x)$ with the rate-of-profit frequency histogram of Fig. 11 and shows 
a reasonable fit between the derived distribution and the data.

\begin{figure}[h]
\begin{center}
\epsfig{file=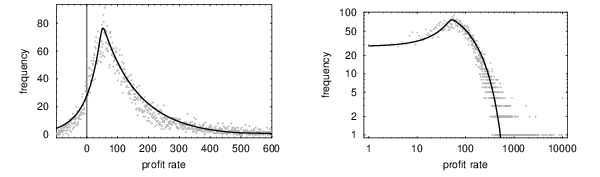,viewport=0 0 300 100,width=1.0\textwidth,clip=true,silent=}
\caption{Theoretical fit to the firm-weighted rate-of-profit distribution. The solid lines plot the theoretical pdf $f_{P}(x)$ scaled by a constant in the frequency axis. The RHS graph plots the function and data in log-log scale and extends the range of the plot to the super-profit range. All profits in excess of 10000 are truncated, which accounts for the outlier at the maximum profit rate.}
\end{center}
\end{figure}

With some further work the 6-parameter distribution $f_{P}(x)$
could be fitted to empirical rate-of-profit measures and compared
against other candidate functional forms. Although (11) ignores effects
due to capital investment the interpretation of the parameters
$\mu_{1}$, $\sigma_{1}^{2}$, $\mu_{2}$ and $\sigma_{2}^{2}$ can be
extended to refer to the means and variances of revenue and cost per
employee. A testable consequence of the SR model is the conjecture 
that the empirical rate-of-profit distribution will be consistent with 
a parameter-mix of a ratio of normal variates with means and variances 
that depend on a firm size parameter that is distributed according to a 
power law.

\section{Discussion}

The empirical coverage of the SR model is broad although the model
can be formally stated in a small number of simple economic rules
that control the dynamics. The model compresses and
connects a large number of empirical facts within a single causal
framework. But this paper only introduces the model and is 
merely a starting point for further analysis and investigation.

The enormous benefit of exploring computational
models of phenomena is that the complex dynamic consequences
of a set of causal rules can be automatically and correctly
deduced by running a computer program that performs a computational
deduction. In this case the deduction is from micro-economic social
relations to emergent, macro-economic phenomena. But the reasons 
why a set of causal rules necessarily generate the observed dynamic 
consequences may initially be opaque precisely because a
computer simulation is required to perform the deduction. This is
why computational modelling is not an alternative to mathematical
modelling but is intimately connected to it. To give just one example,
within the parameter space explored, the SR model generates fluctuations
in national income about long-term stable means. But it requires a
mathematical deduction to understand why this necessarily occurs.
The computational model unequivocally demonstrates that in principle
such a deduction exists and its basic elements and assumptions
will correspond to those of the computational model. Of course, a 
deductive proof may be more or less difficult to construct even
if it is known to exist. So one use of computational modelling
is to more easily identify candidate theories, which may then be 
further analysed to generate explanations in the form of 
mathematical deductions or natural language explanations,
the aim being to understand why the dynamic consequences are logically 
necessary. An example of the potential of this approach is the
deduction of a candidate functional form for the distribution of
industrial profit. Contrast this situation to a purely 
mathematical approach, in which the investigator may only explore 
candidate theories that are directly amenable to mathematical deduction. 
This methodology is 
unnecessarily restrictive, particularly if the system presents 
difficult analytic challenges. 

The model has something to say about a broad range of macro-economic
phenomena that have already been intensively studied and theorised
in standard economic theories, for example business cycle phenomena.
The relationship between the model presented here and existing economic
models of particular economic phenomena is a topic for further research. 
It is likely, for instance, that some of the stochastic models 
elaborated in the econophysics literature may be embedded within the 
overall dynamics of the SR model, particularly those that are more
narrowly focussed on explaining certain distributions in isolation,
such as firm size, income and company growth. However, a new
requirement for more narrowly focussed models is they provide 
more detailed and exact explanations for particular phenomena than 
those provided by the broad but shallow SR model. 

The fact that the empirical distributions considered can be 
deduced from the social relations of production alone suggests that 
some of the striking phenomena of a capitalist economy depend not
so much on specifics but on very general and highly abstract
structural features of that system. In consequence, existing theories 
may be looking in the wrong place for economic explanations, or at 
least introducing redundant considerations. Given this possibility, it 
is worth making a few comments to contrast the approach taken in 
this paper to standard approaches, if only to emphasise that this 
new approach is theoretically motivated.

The ontology of this model differs from standard economic models.
Standard competitive equilibrium models, or neoclassical models, normally take
as their starting point an ontology of rational actors that maximise self-interest
in a market for scarce resources \cite{debreu}. Attention is focussed on determining
the equilibrium exchange ratios of commodity types, which are solutions to a set
of simultaneous, static constraints. Historical time is absent, so
equilibrium states are logically rather than causally derived, and typically
money is not modelled. Neo-Ricardian models, in contrast,
take as their starting point an ontology of technical production relations
between commodity types that define the available material transformations
that economic actors may perform. The production of commodities by means of
commodities \cite{sraffa} results in a surplus product that is distributed to capitalists
and workers \cite{pasinetti}. Despite many essential differences, there are some important
similarities between neo-classical and neo-Ricardian models. For example, prices in
neo-Ricardian models are also exchange ratios determined by solutions to
static, simultaneous constraints. Similarly, historical
time is absent, so there is no causal explanation of how or why a particular
configuration of the economy arose. Money only plays a nominal not a causal
role. There are clear differences between, on the one hand, neo-classical and
neo-Ricardian ontologies, and, on the other,
the basic ontology of the model developed here. Most obvious is that
commodity types and rational actors are absent. Instead, the model emphasises
precisely those elements of economic reality that neo-classical and neo-Ricardian
theories tend to ignore, specifically actor-to-actor relations mediated
by money, which unfold in historical time, and result in dynamic, not static,
equilibria. At a high level of abstraction, and at the risk of over simplification,
neo-classical models theorise scarcity constraints, neo-Ricardian models
theorise technical-production constraints, whereas this model theorises
the dynamic consequences of social constraints, which are historically
contingent facts about the way in which economic production is socially
organised. 

As argued elsewhere \cite{farjoun,lawson} exclusive emphasis on the ontology of
standard economic models is inimical to further progress in the
field of political economy. The SR model constitutes constructive
proof that the standard ontology is redundant for forming explanations
of the empirical phenomena surveyed in this paper. This is not to
assert that some other, perhaps more concrete issues, may require
consideration of purposive activity for their explanation and hence
the introduction of rational actors, or require consideration of
technical production constraints and hence the introduction of commodity
types. Rather, the claim is that, for the empirical aggregates considered, 
there is no need to perform the standard reduction of political economy 
to psychology and the technical conditions of production, and further, that
the dominant causal factors at work are not to be found at the level of
individual behaviour, nor are they to be found at the level of technical-production
constraints, but are found at the level of the social
relations of production, which constitute an abstract, but nevertheless real,
social architecture that constrains the possible actions that purposive
individuals may choose between, whether optimally or otherwise.
This is why the actors in this model probabilistically choose between
possible economic actions constrained only by their class status and
current money endowments, an approach that is closer to the classical
conception of political economy, in which individuals
are considered to be representatives of economic classes that have
definite relations to each other in the process of production.
The social architecture, in particular the wage-capital
social relation, dominates individuals, who, although free to make
local economic decisions, do so in a social environment neither of their
own choosing or control. 

It may be objected that economic actors are clearly purposive
and it is therefore essential to model individual rationality, even 
when considering macro-level phenomena. The underlying assumption of the 
rational actor approach to economics is that macro phenomena are 
reducible to and determined by the mechanisms of individual rationality. 
Farjoun and Machover \cite{farjoun} noted some time ago that the successful
physical theory of statistical mechanics is in direct contradiction to 
this assumption. For example, classical statistical mechanics models
the molecules of a gas as idealised, perfectly elastic billiard balls.
This is of course a gross oversimplification of a molecule's structure
and how it interacts with other molecules. Yet statistical mechanics
can deduce empirically valid macro-phenomena. Quoting Khinchin
\cite{khinchin}:
\begin{quotation}
Those general laws of mechanics which are used in statistical mechanics
are necessary for any motions of material particles, no matter what are
the forces causing such motions. It is a complete abstraction from the
nature of these forces, that gives to statistical mechanics its specific
features and contributes to its deductions all the necessary flexibility.
... the specific character of the systems studied in statistical mechanics
consists mainly in the enormous number of degrees of freedom which these
systems possess. Methodologically this means that the standpoint of
statistical mechanics is determined not by the mechanical nature, but
by the particle structure of matter. It almost seems as if the purpose
of statistical mechanics is to observe how far reaching are the deductions
made on the basis of the atomic structure of matter, irrespective of the
nature of these atoms and the laws of their interaction. (Eng. trans.
Dover, 1949, pp. 8--9).
\end{quotation}
The method of abstracting from the mechanics of individual rationality, 
and instead emphasising the particle nature of individuals, is valid because 
the number of degrees of freedom of economic reality is very large.
This allows individual rationality to be modelled as a highly
simplified stochastic selection from possibilities determined by an
overriding social architecture. The quasi-psychological motives that
supposedly drive individual actors in the rational actor approach
can be ignored because in a large ensemble of such individuals they
hardly matter.

\section{Conclusion}

The aim was to understand the possible economic consequences of 
the social relations of production considered in isolation and 
develop a model that included money and historical time as 
essential elements. The theoretical motivation for the approach is
grounded in Marx's distinction between the invariant social relations of
production and the varying forces of production. Standard
economic models typically do not pursue this distinction.
The model of the social relations of production replicates
some important empirical features of modern capitalism, such as (i) the
tendency toward capital concentration resulting in a highly unequal income
distribution characterised by a lognormal distribution with
a Pareto tail, (ii) the Zipf or power-law distribution of firm
sizes, (iii) the Laplace distribution of firm size and GDP growth,
(iv) the exponential distribution of recession durations, (v)
the lognormal distribution of firm demises, and (vi) the 
gamma-like rate-of-profit distribution. Also, the model
naturally generates groups of capitalists, workers and
unemployed in realistic proportions, and business cycle phenomena,
including fluctuating wage and profit shares in national income.
The good qualitative and in many cases quantitative fit between model
and empirical phenomena suggests that the theory presented here captures
some essential features of capitalist economies,
demonstrates the causal importance of the social relations of production,
and provides a basis for more concrete and elaborated models. 
A testable conjecture is that measures of the empirical 
rate-of-profit distribution will be consistent with a parameter-mix
of a ratio of normal variates with means and variances that depend
on a firm size parameter that is distributed as a power-law.

A final and important implication is that the computational deduction
outlined in this paper implies that some of the features 
of economic reality that cause political conflict, such as extreme 
income inequality and recessions, are necessary consequences of 
the social relations of production and hence enduring and essential
properties of capitalism, rather than accidental, exogenous or
transitory.



\bibliography{ian}
\bibliographystyle{plain}











\end{document}